\documentclass[a4paper,fleqn]{etc/cas-dc}  % Comment this line out if you need a4paper
                                                          
\input{sections/0_packages}

\begin{document}

% --------------------------------------------------------------------
% arXiv Header
% --------------------------------------------------------------------

\twocolumn[
  \begin{@twocolumnfalse}
    \begin{center}{\LARGE\bfseries Preprint Notice\par}\end{center}
    \vspace{2ex}
    \noindent This work has been submitted to the Elsevier 
Reliability Engineering \& System Safety for possible
    publication. Copyright may be transferred without notice, after which this
    version may no longer be accessible.\par
    \vspace{2ex}
    \noindent Submitted to \emph{
Elsevier Reliability Engineering \& System Safety}.\par
    \vspace{2ex}
    \noindent\textbf{Cite as:}\par
    \vspace{1ex}
    \noindent\fbox{\parbox{\dimexpr\linewidth-2\fboxsep-2\fboxrule\relax}{%
     L.~J.~Brettin and N. F. Salem and O. Hans and M. Maurer, ``Teleoperation Operational Design Domain based on Minimal Risk Maneuver Capability'' \emph{
Reliability Engineering \& System Safety}, 2026, submitted for publication.}}\par
    \vspace{2ex}

  \end{@twocolumnfalse}
]
\setcounter{page}{0}
% --------------------------------------------------------------------
% Ende arXiv Header 
% --------------------------------------------------------------------

\shorttitle{Teleoperation Operational Design Domain based on Minimal Risk Maneuver Capability}
\shortauthors{L.\,J.~Brettin et~al.}

\title[mode=title]{Teleoperation Operational Design Domain based on Minimal Risk Maneuver Capability}

\author[1]{Leon Johann Brettin}[orcid=0009-0007-2796-0860]
\cormark[1]
\ead{l.brettin@tu-braunschweig.de}

\author[1]{Nayel Fabian Salem}[orcid=0000-0003-4909-9403]
\ead{n.salem@tu-braunschweig.de}

\author[2]{Ole Hans}[orcid=0009-0009-0239-9469]
\ead{ole.hans@vay.io}

\author[1]{Markus Maurer}[orcid=0000-0002-5357-9701]
\ead{markus.maurer@tu-braunschweig.de}

\affiliation[1]{organization={Institute of Control Engineering, Technische Universit\"at Braunschweig},
            city={Braunschweig},
            postcode={38106},
            country={Germany}}

\affiliation[2]{organization={Department of Operational Safety and Compliance, Vay Technology GmbH},
            city={Berlin},
            postcode={12099},
            country={Germany}}

\cortext[cor1]{Corresponding author}

\begin{abstract}

This article discusses the concept of an Operational Design Domain (ODD) designed specifically for teleoperated road vehicles.
For this purpose, the ODD concept designed for automated driving is adapted for teleoperation.
As teleoperation becomes more common in regular traffic, the question arises under which operating conditions such vehicles are able and allowed to drive.
Currently, these conditions are selected primarily based on network performance.
From a safety perspective, it is difficult to base such a selection on a reliable connection because it is almost impossible to guarantee sufficient reliability.
With this in mind, the ODD concept designed for automated driving is adapted for teleoperation:
A concept is proposed for basing the ODD for a teleoperation system on the capability of the teleoperated vehicle to perform a minimal risk maneuver using a dedicated system designed solely for this purpose.
This concept is then demonstrated using a use case example.
\end{abstract}


\maketitle

\section{Introduction}

The use of teleoperated road vehicles is an emerging field in research and industry.
Two applications are currently prevalent:
First, vehicles entirely driven by a human remote operator.
Second, vehicles that are equipped with an Automated Driving System (ADS) and a teleoperation system as a means of assisting the ADS.

% --- What is the problem? 
For example, a remote operator can be asked to oversee the ADS or to take over the dynamic driving task (DDT) when the ADS cannot continue due to Operational Design Domain (ODD) limitations.
In those cases, it is necessary to ensure the availability and reliability of the remote connection.
However, executing a minimal risk maneuver in such situations can be technically challenging, as the following discussion shows.

% --- Why is it a problem?
\begin{figure}[tbp]
\centering
\includegraphics[width=0.99\columnwidth, page=7]{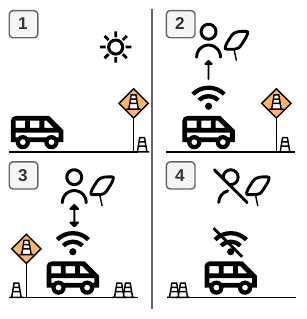}
\caption{
Visualization of a common use case example:
First, the vehicle is driven by an Automated Driving Function (ADS) until it reaches a construction site.
Then, in step 2, the ADS calls a remote operator for guidance.
In step 3, the remote operator guides the vehicle through the construction zone.
Finally, in step 4, the drive through the construction zone is finished, and the remote operator gives control back to the ADS.
This case could happen, both in an urban environment or on a highway, depending on the ODD of the system.
}
\label{fig: common_example}
\end{figure}

The in-the-first-hand practical solution of using teleoperation when the ADS reaches its ODD limits should, from a safety aspect, be seen as critical.
An example of this use case is shown in \cref{fig: common_example}, when the ADS reaches a construction site: 
In step 1, the automated vehicle reaches the construction site and calls in step 2 a remote operator for guidance, as it is unable to drive through the ODD of the construction site independently. 
In step 3 the vehicle system then drives through , assisted by the remote operator, until it reaches the end.
The connection closes in step 4, after which the automated vehicle can continue on its own. 

But what happens in step 3 if the connection degrades and the operator cannot control the vehicle properly while teleoperation is used outside the ODD of the ADS?
If such a degradation is not considered, the system could end up in an undefined state because the ADS cannot execute a proper, minimal risk maneuver because it is outside its ODD.
And the teleoperation system cannot execute a proper maneuver because the connection is degraded.

From a safety perspective, landing in an undefined state is safety critical because such behavior poses a risk of causing harm.
For example, if the vehicle loses connection in step 3, it could collide with construction workers or other vehicles, or it could stop with maximum deceleration, potentially leading to a rear-end crash.
This is critical in an urban scenario, but can be even more hazardous when considering highway scenarios.
All of these scenarios must be considered when the behavior is undefined.
Such a vehicle would most likely not pass any safety requirements, making it impossible to be approved for release without a safety driver.

% --- What is the solution?
To address this issue, we proposes a systematic definition of a Teleoperation Operational Design Domain in combination with an ADS to ensure that it does stay in a defined state, capable of executing a minimal risk maneuver.
To achieve this, the capability to reach an adequate minimal risk condition serves as the criterion delimiting the ODD of the vehicle system, for both ADS operation and the teleoperation system.

% --- Why does the solution work
The problem of how to handle the connection as a safety-critical component can be shifted to ensuring the minimal risk maneuver (MRM) functionality.
This solves the ambiguity of how to handle the connection part in a safety argumentation.

Handling the problem on relying on the connection for teleoperation systems is, as seen by e.g., BSI Flex 1886~\cite{BSI1886_2023} and the German Federal Highway Research Institute~\cite{WG_Teleoperation2024_en}, still an ongoing research question we want to address here. 

The article is structured as follows:
\cref{sec: terminology} defines the terminology used in this article. 
\cref{sec: related work} discusses work related to this article.
\cref{sec: discussion} presents a discussion of the handling of, and ambiguities in, the ODD of the ADS. 
The proposed handling of an ODD with teleoperation in mind is shown in \cref{sec: main idea}.
A use case example is given in \cref{sec: use case example}.
Finally, the conclusion is given in \cref{sec: conclusion}.

\section{Terminology and Definitions}
\label{sec: terminology}

This section provides definitions of the important terms used throughout the article.

\subsection{Automated Driving System (ADS)}

The \textit{Automated Driving System} (ADS) is defined in SAE J3016 \cite{J3016} as ``[t]he hardware and software that are collectively capable of performing the entire DDT on a sustained basis, regardless of whether it is limited to a specific ODD; this term is used specifically to describe a Level 3, 4, or 5 ADS''~\cite[Clause 3.2]{J3016}.
This includes, for example, the capability to control the vehicle, identify the surroundings of the vehicle through specific sensors, and act accordingly. 
Control of the vehicle with the ADS is limited for Level 3 and 4 vehicles to a specific ODD, which is discussed in \cref{sec: terminology ODD}.

\subsection{Teleoperation System}

\begin{figure}[tbp]
\centering
\includegraphics[width=\columnwidth, page=6]{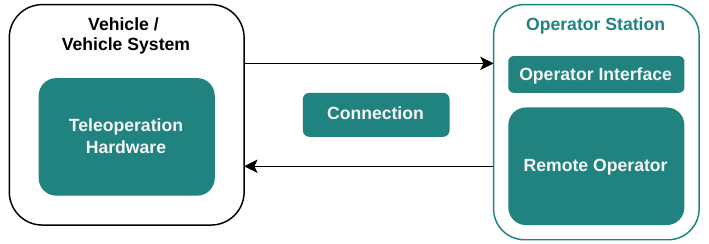}
\caption{
    The basic architecture of a teleoperation system. The architecture is based on the concept of \textcite{Winfield2009}.
}
\label{fig: teleoperation_system}
\end{figure}

According to \textcite{Winfield2009}, a teleoperation system consists of three components: 
\begin{enumerate}
  \item Operator Interface
  \item Communication Link
  \item Robot
\end{enumerate}

For the purposes of this article, the terms \textit{operator station} and \textit{remote operator} are used to refer to the location from which the robot is controlled and the person who controls it.
The communication link is called \textit{connection}, and because this article focuses on road vehicles, the term \textit{vehicle} will be used instead of the broader term \textit{robot}.
For the purposes of this article, the \textit{connection} is a radio link that sends signals between the vehicle and the operator station over a wireless network (e.g. 5G).
The entire system responsible for executing the DDT [Dynamic Driving Task], including the ADS and the teleoperation system, will be referred to as the \textit{vehicle system} in this article.
An overview of the terminology used can be seen in \Cref{fig: teleoperation_system}.

Two modes are commonly used to describe the mode of operation for teleoperation: \ra and \rd .
There are other variants to these modes that are discussed by \textcite{Majstorovic2022} but are beyond the scope of this article.

\subsubsection{Remote Assistance}
\label{sec: terminology Remote Assistance}

\ra is defined in the J3016 as ``[e]vent-driven provision, by a remotely located human [...], of information or advice to an ADS-equipped vehicle in driverless operation in order to facilitate trip continuation when the ADS encounters a situation it cannot manage''~\cite[Clause 3.23]{J3016}. 

It is important to note that \ra can only be used if the vehicle is equipped with an ADS.
Also, as stated, \ra is used ``when the ADS encounters a situation it cannot manage'', which will be discussed in this article, because we show that a non manageable situation could lead to unreasonable risk.

Looking at current press or media presentations of this mode\footnote{e.g., from Zoox \url{https://zoox.com/journal/humans-in-the-loop} (status as of 21 May 2026) or Waymo~\cite{Waymo2024_Fleet_Response}}, mostly the ADS or a system similar to an ADS is still active on the vehicle and with the help of waypoints or other actions that can be done via mouse and keyboard, the system is supported by the remote operator. 
This system still performs the lower-order driving functions, following the advice of the \ro .

\subsubsection{Remote Driving}
\label{sec: terminology Remote Driving}
Remote driving, on the other hand, is defined in SAE J3016 as ``[r]eal-time performance of part or all of the DDT and/or DDT fallback (including, real-time braking, steering, acceleration, and transmission shifting), by a remote driver''~\cite[Clause 3.24]{J3016}. 

Remote driving does not necessarily require an ADS installed on the vehicle. 
To give a simple example: Most often, \rd can be thought of as an \ro controlling the vehicle via a steering wheel and pedals from an operator station\footnote{e.g., from Elmo \url{https://www.elmoremote.com/} or Vay~\cite{Hans2024_our_teledriving}}.

\subsection{Operational Design Domain}
\label{sec: terminology ODD}

The \textit{Operational Design Domain (ODD)} is defined as ``[o]perating conditions under which a given driving automation system or feature thereof is specifically designed to function, including, but not limited to, environmental, geographical, and time-of-day restrictions, and/or the requisite presence or absence of certain traffic or roadway characteristics''~\cite[Clause~3.21]{J3016}.

In this article, we also use this definition for teleoperation systems, although it was not designed for teleoperation systems.
The argumentation for using the term \textit{ODD} here is that we will discuss a necessity of a limitation of the teleoperation system based on the capability of the system, which can be mapped to the term ODD for autonomous driving.
The use of the term ODD for teleoperation can also be found in other publications, e.g., \cite{BSI1886_2023, Hans2025_Evaluation, Hans2025_Identification}.

\subsubsection*{ODD Border}
ISO~34503~\cite{ISO34503} uses the term \textit{ODD Border} but does not directly define it.
In short, it can be described as a set of one or multiple combined attributes that define the limits of the ODD. 
When the ODD border is reached, according to \cite{ISO34503}, a minimal risk maneuver or transition to a fallback-ready user shall be initiated. 

It is important to note that when an ODD border is reached and an ODD transition is initiated, the system transitions from \textit{inside} the ODD to \textit{outside} the ODD.

\subsection{Minimal Risk Condition / Minimal Risk Maneuver}
A minimal risk \textit{condition} is defined in SAE J3016 as ``[a] stable, stopped condition to which a user or an ADS may bring a vehicle after performing the DDT fallback in order to reduce the risk of a crash when a given trip cannot or should not be continued''~\cite[Clause~3.16]{J3016}.

A minimal risk \textit{maneuver} on the other hand is defined as ``[...] a manoeuvre aimed at minimizing risk in traffic by stopping the vehicle in a safe condition (i.e. minimal risk conditions)''~\cite[Article 2, Clause 14]{EU2022_1426} or as defined in ISO/TR~4804 as ``automated driving system’s [...] capability [...] of transitioning the vehicle between nominal and minimal risk conditions [...]''~\cite[Clause 3.30]{ISOTR4804:2020}.

In this article, we use the term \textit{minimal risk condition} instead of \textit{safe condition} because the definition of ``safe'' may depend on the audience or the understanding of society (see \citeauthor{Reschka2016_safety}~\cite{Reschka2016_safety}, \citeauthor{Salem2024_safety}~\cite{Salem2024_safety} and \citeauthor{Fleischer2023_Oberseminar_eng}~\cite{Fleischer2023_Oberseminar_eng}).
And, as stated by \textcite{Stolte2022}, a minimal risk condition that exceeds the accepted risk threshold would not qualify as a \textit{safe state}.
``\textit{safe state}'' is defined there as ``[s]tate in which a system does not pose an unreasonable risk''~\cite{Stolte2022}. 
However, the overall system has to be safe even when applying an MRM.

\subsubsection*{A closer look at SAE J3016}

SAE J3016 does not directly address the case of using \rd or \ra outside of the ODD of an ADS.
It only states that an ADS can be used briefly somehow outside its ODD (which, according to SAE J3016 does not necessarily represent an ``ODD exit'') if a rapid environmental change occurs, such as ``inclement weather'' or ``obscured lane lines''.
Otherwise, SAE J3016 provides the requirement for the ADS to perform a fallback when moving outside the ODD.
%See:

%\begin{displayquote}
%``NOTE 6: While performing DDT fallback, an ADS may operate temporarily outside of its ODD (see 3.21 NOTE 1).''
%\cite[Clause~3.12 Note~6]{J3016}
%\end{displayquote}
%\vspace{1em}
%\begin{displayquote}
%``NOTE 1: While Level 3 and 4 ADS features/vehicles are designed to operate exclusively within their respective ODDs,  some ODD conditions are subject to rapid change during on-road operation (e.g., inclement weather, obscured lane lines). Such transient changes in the operating environment do not necessarily represent an 'ODD exit,' as the ADS determines when such a change in conditions requires fallback performance (whether by the fallbackready user or ADS).''\\\cite[Clause~3.21 Note~1]{J3016}
%\end{displayquote}

But if it is planned to use teleoperation outside of the ODD anyways, this can technically be done either to support the ADS by giving it "information or advice"~\cite[Clause~3.23]{J3016} through \ra , or to take over the dynamic driving task completely through \rd .

%\begin{displayquote}
%\textit{Definition of \ra}:\\
%``Event-driven provision, by a remotely located human (see 3.31.5), of information or advice to an ADS-equipped vehicle in driverless operation in order to facilitate trip continuation when the ADS encounters a situation it cannot manage.''\\
%\cite[Clause~3.23]{J3016}
%\end{displayquote}

Either way, the vehicle still needs some kind of automatic fallback mechanism when the connection is lost.
And with an ADS on board it potentially has the option of using the ADS, or parts of it, for the minimal risk maneuver.
However, depending on how the ODD for the ADS is defined, relying on the ADS for such a maneuver can be quite problematic.

%----------------------------------------------------------------------------------------------------------------
SAE J3016 states for a \textit{dynamic driving task fallback} that it will either be executed (1) when a performance-relevant system failure occurs, or (2) upon ODD exit~\cite[Clause~3.12]{J3016}.
%\begin{displayquote}
%\textit{Definition of Dynamic Driving Task}:\\
%``The response by the user to either perform the DDT or achieve a minimal risk condition (1) after occurrence of a DDT performance-relevant system failure(s), or (2) upon operational design domain (ODD) exit, or the response by an ADS to achieve minimal risk condition, given the same circumstances.''
%\cite[Clause~3.12]{J3016}
%\end{displayquote}

However, it does not address either of those parts for teleoperation usage out of the ODD of the ADS.
For the first point, a disconnect or a degradation of the connection quality below a sufficient quality can be classified as such a failure, as this results in the loss of controllability for the remote driver.
But SAE J3016 exempts this from its consideration (see~\cite[Clause~8.5.iii]{J3016}) and refers to
a failure mitigation strategy (see \cite[Clause~3.11]{J3016}).
%\begin{displayquote}
%``Fallback performance and minimal risk condition achievement at Levels 4 and 5 require that the ADS is still functional  after occurrence of a DDT performance-relevant system failure or out-of-ODD condition. If the ADS is not functional, a failure mitigation strategy may apply (see 3.11 and 8.6). The minimal risk condition depends on both the vehicle condition and its operating environment at the time that fallback is triggered and could follow a degraded mode strategy that considers the relative risks associated with continuing operation, pulling off the road, or stopping in place.''\\
%\cite[Clause~8.5.iii]{J3016}
%\end{displayquote}
%\begin{displayquote}
%\textit{Definition of Failure Mitigation Strategy}\\
%``A vehicle function (not an ADS function) designed to automatically bring an ADS-equipped vehicle to a controlled stop in path following either: (1) prolonged failure of the fallback-ready user of a Level 3 ADS feature to perform the fallback after the ADS has issued a request to intervene, or (2) occurrence of a system failure or external event so catastrophic that it incapacitates the ADS, which can no longer perform vehicle motion control in order to perform the fallback and achieve a minimal risk condition. (See 8.6.)''\\
%\cite[Clause~3.11]{J3016}
%\end{displayquote}
Stopping in lane is an example given by SAE J3016 \cite[Figure~13]{J3016} for such a strategy, but it is also stated that failure mitigation is different from achieving a minimal risk condition and is out of scope for the taxonomy described in SAE J3016 \cite[Clause~8.6]{J3016}.

For the \mbox{second} point, when a connection degrades to a non-usable level during teleoperation (when teleoperation is used as a backup for an ADS) then there is no transition from an operation condition in which the ADS is designed to function, because the system is already out of its conditions.

\subsubsection*{Minimal Risk Maneuver can be unsafe}

What must be made clear is that a maneuver that  the vehicle on the road without analyzing the environment cannot be called a minimal risk maneuver:
It \textit{can} end in a minimal risk condition, but so can any other maneuver.
A minimal risk maneuver shall \textit{aim} at reaching a minimal risk condition, but it might not be sufficient to ensure safe behavior.
To minimize the risk, there needs to be a consideration of the current environment and a calculation of the potential risk.
For example, the vehicle could be equipped with the capability to perceive its environment and decide on a trajectory that would minimize risk.
It is worth noting that simply stopping or executing a minimal risk maneuver does not necessarily result in an acceptable level of risk.
This understanding of a minimal risk maneuver, and the ambiguity that a minimal risk maneuver can be unsafe, is a research topic on its own, which, given the focus of this article, we cannot go into in more detail here.
This research topic can even be of interest for current regulation and has potential for future work referencing this article.

\subsubsection*{Which actor is used to achieve a minimal risk condition?}
Different sources identify different actors that can realize the minimal risk maneuver:

\begin{itemize}
    \item According to J3016~\cite[Clause 3.16]{J3016}, the condition is reached by a user or an ADS. 
    \item ISO~21448~\cite[Clause~3.16]{ISO21448} does not specify an actor in its definition of a minimal risk condition.
    \item ISO/TR~4804~\cite[Clause~5.2.6]{ISOTR4804:2020} states that the a minimal risk maneuver ``is the system's capability of transitioning the vehicle between minimal risk conditions'', making the ADS the actor.
    \item \textcite{Ackermann2023}, uses a system separate from the ADS to perform the minimal risk maneuver. 
\end{itemize}

Depending on the definition, these actors are not completely different. 
For this article, we will define the actor that initiates the minimal risk condition as a system \mbox{capable} of doing so.
It can be the ADS, but it can also be another dedicated subsystem, similar to ISO/TR~4804 and the implemented version of \textcite{Ackermann2023},

\subsection{Failure}
\textcite{Stolte2022} introduce a taxonomy to unify fault tolerance regimes for automotive systems, providing consistent definitions for terms in this research area.
In this article, the terms \textit{operational}, \textit{fail-degraded}, and \textit{fail-safe} are particularly important, because they provide a clear definition of how to define failure of the connection between the remote driven vehicle and the operator station.

\textit{Operational} is defined as follows:
``An operational system has no fault and, thus, can provide its specified functionality with at least nominal performance while maintaining a safe state''~\cite{Stolte2022}.

It is important to note that \citeauthor{Stolte2022} differentiate between a safe state and a minimal risk condition.
According to the authors, the minimal risk condition is reached by the vehicle in response to specific events due to the implemented safety mechanism.
However, this condition ``may exceed the accepted risk threshold'' needed to be considered a \textit{safe state}, which is defined as a ``[s]tate in which a system does not pose an unreasonable risk''~\cite{Stolte2022}.

A system is \textit{fail-safe} when a fault combination is present, which ceases its specified functionality and [the system] transitions to a well-defined condition to maintain a safe state''~\cite{Stolte2022} 
and \textit{fail-unsafe}, when the system is ``in the presence of a fault combination if it is not able to maintain a safe state''~\cite{Stolte2022}.

A system is \textit{fail-degraded} when a fault combination is present and the system still ``can provide its specified functionality with below nominal performance while maintaining a safe state''~\cite{Stolte2022}.

In section \ref{sec: main idea}, these definitions are used to describe the different states in which the driving automation system can be, depending on the implementation of the capability to execute a minimal risk maneuver.

\subsection{System of Systems / Systems Engineering}

\begin{figure*}[tbp]
\centering
\includegraphics[width=0.99\textwidth, page=10]{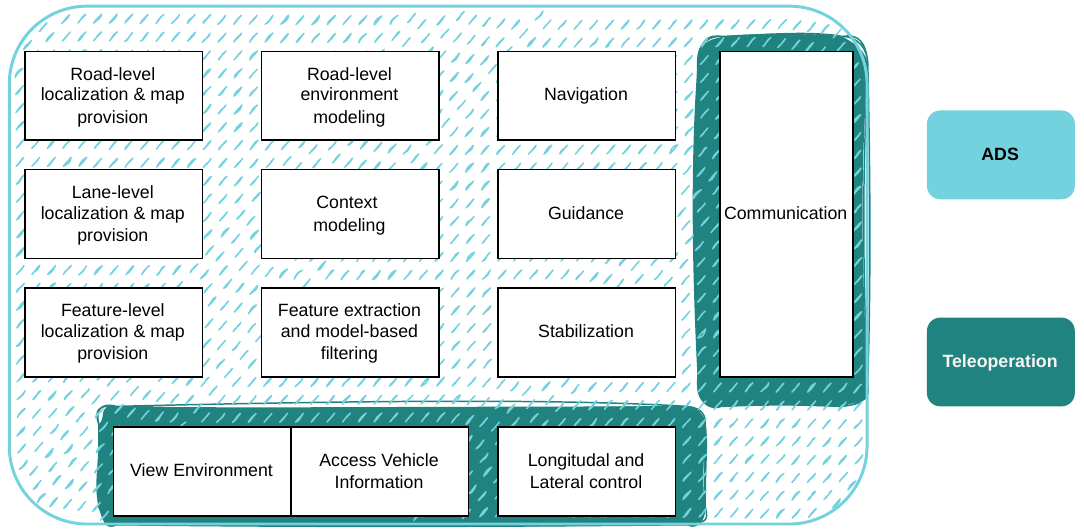}
\caption{
Functional components of an autonomous vehicle can be viewed as its capabilities.
The components shown are based on \textcite{Ulbrich2017}, with adjustments made for this article.
Some vehicle system functional blocks are used by both the ADS and the teleoperation system.
This is a capability-based view. Even when using an end-to-end model for this problem, these capabilities must still be implicitly incorporated into the system.
}
\label{fig: capabilities}
\end{figure*}

In the context of systems engineering and considering the interoperability between an ADS and a teleoperation system, the entire vehicle system can be viewed as a \textit{system of systems}.
This overall system has multiple capabilities, such as longitudinal and lateral control of the vehicle, vehicle navigation, and vehicle information access.
The capabilities can be exhibited by subsystems like the ADS and the teleoperation system.
Most system capabilities can be further divided into finer-grained capabilities, as seen in \cite{Reschka2017, ISO19540, Nolte2025}. 
An example of this can be seen in \cref{fig: capabilities}.

This article mainly uses the system of systems definition to demonstrate that the ADS and the teleoperation system share a connection capability, but have different criticalities, as discussed in \cref{sec: discussion}. 

\subsection{Controllability}

The definition of controllability varies in different research areas. The two areas that come closest to the subject matter of this article are control engineering and ergonomics.
Controllability in control engineering terms is e.g., defined ``as the ability to reach the origin and an arbitrary state, respectively''~\cite{Astrom2012} after \cite{Kalman1969}.
In ergonomics it is defined in ISO 92441 as follows:
"Controllability: The interactive system allows the user to maintain control of the user interface and the interactions, including the speed and sequence and individualization of the user-system interaction."\cite[Clause 4.1]{ISO9241-2020}. 

For the purposes of this article, losing controllability means that it is no longer possible to interact with the vehicle in order to reach a desired location.

\section{Related Work}
\label{sec: related work}

% ---------------------------------------------
% Introduction to Related Work
% ---------------------------------------------
The combination of teleoperation and ODD considerations is not an entirely new field.
In teleoperation, the problem of maintaining a reliable connection is implicitly and explicitly related to such considerations, as will be demonstrated in this section.
The focus on the connection is evident in decisions about where to drive, depending on factors such as latency, jitter, and connection availability. These factors limit the operating area to an implicit geographic ODD.

First, we will present work that considers the ODD in this manner (\cref{sec: related work -- combination}).
Because the term ODD is defined for automated driving, the next subsection (\cref{sec: related work -- formalization}) will present research that considers ODD formalization.

As this article aims to demonstrate the integration of teleoperation and connection degradation, the next subsection (\cref{sec: related work -- degraded}) will present research from the automated driving domain that considers the consequences of degradation on the ODD.

% ---------------------------------------------
% Safety Blind Spots (Brettin2025)
% ---------------------------------------------
Before presenting the related work, in a previous work \cite{Brettin2025} by several of the authors of this article, the question was raised about what would happen if the connection to a remotely driven vehicle were lost, causing it to stop in a straight line on an urban road with a vehicle following from behind.
To answer this question, different braking deceleration values and reaction times of following vehicles were simulated, and a conceptual risk assessment was performed based on the fallback strategy of straight-line braking.
The result was that, based on the assumptions made in the article, a fallback strategy of straight-line braking can be considered unsafe in this case.
This led the authors to ask what an alternative might look like, which led to the work in this \todd article.

\subsection{ODD in combination with teleoperation}
\label{sec: related work -- combination}

This subsection considers related work, which focuses on the combination of ODD concepts with teleoperation concepts. 
As this work proposes a \todd these sources and their influence on this article have to be mentioned and discussed.

% ---------------------------------------------
% Research Needs Teleoperation (WG_Teleoperation_2024_en)
% ---------------------------------------------
In \cite{WG_Teleoperation2024_en}, research questions for teleoperated vehicles were collected for different clusters. 
Some of the authors of this article were part of the cluster ``Vehicle, area of operation and functional safety''.
One of the questions asked was what would happen if a teleoperated vehicle left the ODD of the ADS and drove in a \todd without using an ADS capable of operating in this ODD.
Here, questions were asked about the ability to reach a minimal risk condition when the teleoperation system is not functioning correctly \cite[Chapter 4.1.5.4]{WG_Teleoperation2024_en}.
This will also be discussed in this article.

% ----------------------------------------
% Fernlenkverordnung
% ----------------------------------------
The German government recently published a regulation permitting remotely driven vehicles in Germany (StVFernLV)~\cite{StVFernLV2025}.
The regulation is highly similar to the German regulation regarding Level 4 vehicles (AFGBV~\cite{AFGBV2022}) but does not explicitly address Level 4 functions or vehicles.
The regulation requires a target Operational Design Domain in which the remotely driven vehicle is permitted to operate.
This operational domain mainly focuses on geographical definitions \cite[Appendix~2,1.4~a-k]{StVFernLV2025},  but also considers network coverage \cite[Appendix~2,1.4~l-m]{StVFernLV2025}, weather, time, and traffic \cite[Appendix~2,2.2-2.3]{StVFernLV2025}.
The regulation also requires a function to execute a minimal risk maneuver but does not specify in detail how this function must be designed.

This article focuses on the combination of an ADS and the use of a teleoperation system (see e.g., \Cref{sec: terminology Remote Assistance} on Remote Assistance) which differentiates it from the aforementioned regulation document, which only considers teleoperation systems without ADS (see e.g., \Cref{sec: terminology Remote Driving} on \rd ).
However, it may aid in developing how such a \todd can be implemented there as well.

%
% ---------------------------------------------
% Vay Framework (Hans2023)
% ---------------------------------------------
An ODD qualification framework is presented in \cite{Hans2023}.
It presents a step-by-step approach to be used when a new region is to be added to the driving environment of a remotely driven vehicle. 
\textcite{Hans2023} mainly focus on regions and connections for their ODD definition:
In a first step, when a new region is added to their ODD, an analysis for regional specifications such as speed limits of roads is done.
Then an analysis of the connection is performed using a ``connectivity failure rate prediction model'', to reduce the collection of a hard-to-achieve amount of data that would be needed to statistically prove a connection loss. 
A performance validation of the connectivity, the system and of the remote operator is performed to check if all are ready for the newly added region. 
If all are ready, the new region is ready for integration. 
During real-world operation, monitoring is performed for potential changes. 
If anything changes, the vehicle is put into a minimal risk state until further evaluation.
The ODD is limited here to regional boundaries, and the focus is on connectivity.

% ---------------------------------------------
% ODD design for automated and remote driving  systems: A path to remotely backed autonomy (Hans2025_ODD)
% ---------------------------------------------
In a subsequent paper, \textcite{Hans2025_ODD} present a comparison between the ODD of an ADS and \rd based on the PEGASUS layers.
The paper compares different strengths and weaknesses of autonomous driving and \rd and concludes that the concepts can potentially complement each other. 
%
%
% ---------------------------------------------
% Remote Assistance or Remote Driving: The Impact of Operational Design Domains on ADS-Supporting Systems Selection
% ---------------------------------------------
This is followed up by a comparison between the usage of \rd and \ra as supporting technologies for an ADS on the basis of the PEGASUS Layers \cite{Hans2025_Impact}.
The paper concludes that, for using these layers, there are different pros and cons for the usage of either \rd or \ra .

% ---------------------------------------------
% PEGASUS Method
% ---------------------------------------------
For the PEGASUS Method, \textcite{Hans2025_ODD,Hans2025_Impact} refer to \textcite{Wachenfeld2016_PEGASUS}, which focuses primarily on the safety assurance of the PEGASUS project and a refinement of the PEGASUS layers by \textcite{Scholtes2021}. 
Regarding the layer models introduced in the PEGASUS-Project~\cite{PEGASUS2020_Abschlussbericht} / PEGASUS-Method~\cite{PEGASUS_Method}, we also recommend the work of \textcite{Schuldt2017} for the initial concept of the layer models, which was further developed by \cite{Bagschik2018_Szenariengenerierung, Bagschik2018_Ontology} with the addition of a fifth layer and \textcite{Bock2018} with the addition of a sixth layer.
In the follow-up project VVM~\cite{VVM_Abschlussbericht}, \textcite{Scholtes2021} further develop the concept of the 6-layer model from the sources mentioned above.

% ---------------------------------------------
% Levels of autonomous driving (Zhang2020) 
% ---------------------------------------------
\textcite{Zhang2020} distinguishes \mbox{between} different levels of human influence: from levels where the human remote operator has to do the whole driving task, to levels with more and more automated functionality, where the human operator has to do less for each level of the automated system.
This work is important for the proposed \todd as it has similarities to the method described in this article.
The proposed \todd system also has some parts where the human operator is driving and some parts where the system must be automated and no human intervention is possible.
To ``ensure safety'', \citeauthor{Zhang2020} also points out the need for (1) \textit{awareness} of a potential loss of connection, (2) \textit{precautions} to enter a safe mode before control is lost, and (3) a \textit{self-survival} function that, depending on the duration of the disconnection, will stop, pull over, or drive the vehicle into a safe mode.
These topics will be addressed in this article, because the need for a \todd based on the vehicle's capability to reach a minimal risk condition targets all three of \citeauthor{Zhang2020}'s points, with a focus on the third point.

% ---------------------------------------------
% Free Corridor (Diermeyer2011, Tang2014, Hoffmann2022)
% ---------------------------------------------
\textcite{Diermeyer2011}, \textcite{Tang2014}, and \textcite{Hoffmann2022} propose the free corridor method.
Here, a corridor in front of the remote-driven vehicle must be kept free, so that if the vehicle loses the connection, the vehicle has a corridor free of obstacles in which to brake.
This also addresses the issue of how a vehicle should behave in the event of a loss of connection, which is also one of the main issues addressed in this article.

% ---------------------------------------------
% Route Planning system (Neumeier2019)
% ---------------------------------------------
\textcite{Neumeier2019} (also mentioned in a previous work~\cite{Neumeier2019_feasability}) propose the idea of a route planning system that only plans routes for remote-driven vehicles with sufficient network performance.
Also, depending on the network performance, speed adjustments of the remote-driven vehicles are applied.
Here, the speed adjustments depend on the distance to the vehicle in front, the maximum velocity based on the braking distance, and a maximum velocity for driving around curves.

The work proposed in this article uses the ODD as a critical factor that limits where a remote-driven vehicle can and cannot drive.
\citeauthor{Neumeier2019} also uses a route planning function to decide where the remote-driven vehicle may drive and where it may not and also limits the maximum speed based on the capabilities of the teleoperation system.
The mentioning of capabilities is similar to this work, too, because the capabilities of the whole system are relevant to decide where and how to drive as seen in e.g. \cref{sec: main idea}.

% ---------------------------------------------
% Risk-Aware Shared Control for Teleoperation of Automated Vehicles in Dynamic Environments (Brecht2024_Risk)
% ---------------------------------------------
Activated when ``the vehicle has reached the boundary of its Operational Design Domain''~\cite{Brecht2024_Risk}, in \textcite{Brecht2024_Risk} the remote operator generates trajectories that are used to calculate a risk metric.
The risk metric is calculated primarily based on ``dynamic objects that may intersect the planned trajectory when entering the ego vehicle's path from the side''~\cite{Brecht2024_Risk}.
Based on this risk metric, a trajectory is selected that slows the vehicle if necessary or uses a deceleration that best fits the given scenario.
The method relies on a working prediction and filtering of objects from the AV stack.

This is similar to the work proposed in this article, because both methods involve the use of teleoperation from its ODD, as well as a type of risk analysis to determine its trajectories.
However, in this work, we do not focus on improvements during teleoperation but rather on the broader picture of overall capability, as will be shown in the next sections.

% ----------------------------------------
% BSI Flex 1886 System aspects for remote operation of vehicles -- Guide (BSI1883_2023)
% ---------------------------------------------
An ODD for teleoperated vehicles is also discussed in BSI~Flex~1886~\cite{BSI1886_2023}, which states that ``[d]efining an ODD specifically for a class of remote operation is a critical part of the safety case''~\cite[Clause 3.6]{BSI1886_2023}.
The Flex Standard discusses a wide range of system aspects of remotely operated vehicles, from use case descriptions to system descriptions to what information a remote operator needs, all from a safety perspective.
In comparison to this work, it is important to note that the standard demands ``on-board self-driving capabilities, even if they are not intended to drive themselves continuously, at least to provide safe execution of an MRM [Minimal Risk Maneuver] when the communication link is degraded or fails''~\cite[Clause 9.2]{BSI1886_2023}.
It is also stated that this problem is still an ``area of research and development'' also mentioning that stopping in a straight line is ``unlikely to be a low-risk manoeuvre in many driving scenarios''~\cite[Clause 9.4]{BSI1886_2023}. 
This is of high importance for this article, as this \todd work directly addresses the ``area of research and development''.

The standard was also developed in combination with BSI~Flex~1887~\cite{BSI1887_2025}, which focuses on aspects regarding the human operator of the remote vehicle.

% ----------------------------------------
% Task Allocation And Control Transitions In Autonomous Driving System Operations (Correa-Julian2024)
% ----------------------------------------
\textcite{CorreaJullian2024_Task_Allocation} propose a taxonomy for control transitions between ADS and different  types of operators, including remote operators.
They explicitly do not focus on connection loss but on mode transitions during normal ADS operation.
This article also considers transitions between ADS and remote operator, but with a focus on connection loss as the trigger for a minimal risk maneuver.

\subsection{Operational Design Domain formalization}
\label{sec: related work -- formalization}

The following related work consider different ways of formalizing an ODD. 
They are mentioned here, because they aim to structure the work with ODD descriptions, including not only geographical limits, but also a broader picture of the ODD.
This is important for this work because it demonstrates that the definition of an ODD is influenced by different factors, and that the connection component is only a part of the ODD.

% ----------------------------------------
% A Two-Level Abstraction ODD Definition Language (Irvine2021) and (Schwalb2021)
% ----------------------------------------
ISO34503~\cite{ISO34503:2023} defines an ODD taxonomy and definition format for Level 3 and Level 4 Automated Driving Systems (as stated in this standard, Level 5 does not require an ODD definition because the ODD of a Level 5 ADS is unlimited).
The ODD format provides manufacturers with a standardized way to define ODD-based capabilities and safety restrictions.
The top-level categories in this standard are: ``scenery elements'', ``environmental conditions'', and ``dynamic elements''.
Among the more specific definitions is one for a \textit{target operational domain}, which describes the operating conditions in which the vehicle is expected to operate. This is in contrast to the ODD, which describes the conditions in which the vehicle is capable of operating.

% ----------------------------------------
% A Two-Level Abstraction ODD Definition Language (Irvine2021) and (Schwalb2021)
% ----------------------------------------
The two papers \cite{Irvine2021, Schwalb2021} present a structured ODD description model.
The first part, by \citeauthor{Irvine2021}, presents a structured textual description format for an ODD.
The second part, by \citeauthor{Schwalb2021}, presents a machine-readable description based on the formalization of the first part.

% ----------------------------------------
% Operational Design Domain for Automated Driving Systems: Taxonomy Definition and Application (Mendiboure2023)
% ----------------------------------------
\textcite{Mendiboure2023} compare different ODD-taxonomies and create a joint taxonomy based on their findings. 
Overall the paper argues that all found taxonomies are related to J3016~\cite{J3016} but differ slightly in the amount of categories they include. 
\citeauthor{Mendiboure2023} then propose the following categories: ``Physical Infrastructure'', ``Scenery'', ``Environmental Conditions'', ``Traffic conditions'', ``Digital Infrastructure'', and ``Vehicle capabilities''.

% ----------------------------------------
% Towards approval of autonomous ship systems by their operational envelope (Rodseth2021 und Rodseth2022) 
% ----------------------------------------
\textcite{Rodseth2021_ODD_OE, Rodseth2022_OE_approval} propose an \textit{operational envelope} for ships. 
The envelope is based on the ODD definition from automated driving, but additionally includes the capabilities of human ship operators and fallbacks. 
This addresses a gap similar to the one discussed in this article.
In contrast, the \todd proposed in this article is bounded by the vehicle system's capability to perform a minimal risk maneuver and thus does not rely on the remote operator or the connection to reach a safe state.

% ----------------------------------------
% Addressing systemic risks in autonomous maritime navigation: A structured STPA and ODD-based methodology (Nakashima2025)
% ---------------------------------------------
\textcite{Nakashima2025} present a structured method based on System-Theoretic Process Analysis (STPA)~\cite{Leveson2018} and ODD concepts to systematically address systemic risks in autonomous maritime navigation. 
They also develop an ODD taxonomy tailored to autonomous ships and their specific needs. 
This article focuses on a single risk, the loss of connection of a teleoperated vehicle, and likewise proposes an ODD-based solution for it.

\subsection{Degraded ADS}
\label{sec: related work -- degraded}

When an ADS is unable to continue its dynamic driving task and a teleoperation system is called in to assist, the ADS is in a degraded state. 
The following related work discusses ADS degradation. While none of these sources directly address teleoperation, the concepts are common in the relevant fields. 
This is why they are mentioned here.

% ---------------------------------------------
%  restricted-ODD (Colwell2018)
% ---------------------------------------------
\textcite{Colwell2018} propose the idea of a \textit{Restricted Operational Domain} and a \textit{Degraded Operation Mode}:
A Restricted Operational Domain is a subset of the ODD under which an automated vehicle is designed to function. 
Depending on the current operation mode of the vehicle, the vehicle may still be able to function, but may have more limited capabilities due to degraded subsystems.
\citeauthor{Colwell2018} give the example of a sensor failure that reduces the vehicle's detection range. 
In this case, the automated vehicle can still operate in its limited operating range, but not in the full ODD for which it was designed. 
The vehicle is operating in a degraded mode because a system or subsystem is degraded, but it can still drive (though in a degraded/limited way) and potentially complete its mission.

% ---------------------------------------------
%  mu-ODD (Koopmann2019)
% ---------------------------------------------
\textcite{Koopman2019} introduces the concept of $\mu$\textit{-ODD's}, where the ODD is divided into several smaller parts. 
The idea behind the concept is that some use cases can be improved by having a specific ODD for them.
An example given is a driving situation that can occur often and has a long duration, a vehicle is driving for a long time at \qty{50}{\kilo\metre\per\hour} on a dry road. 
Another example given is driving on an icy road. 
In this $\mu$-ODD, the driving behavior of the automated vehicle can be set to drive more slowly and more carefully. 
Overall, \citeauthor{Koopman2019}'s concept offers the chance to focus more on specific ODD-related problems.

Both concepts, \citeauthor{Colwell2018} and \citeauthor{Koopman2019}, use a similar base idea as in this article of specifying and limiting the ODD only for more specific use cases, which is similar to explicitly specifying it for the capabilities of a minimal risk maneuver.
The difference is that \citeauthor{Colwell2018} and \citeauthor{Koopman2019} focus on automated vehicles, not teleoperated vehicles. 
Additionally, Restricted Operational Domains and $\mu$-ODDs have broader use cases because they do not specifically address minimal risk maneuvers.
The perspective is also different. In this work, an ODD is defined by the capability to perform a minimal risk maneuver. In Colwell's work, the vehicle can operate with degraded subsystems and still perform its missions. Our focus is on transitioning directly to a minimal risk condition when something malfunctions.
Additionally, when examining vehicle capabilities, the full ODD of the system is defined by the capability of the degraded mode. In contrast, the degraded ODD in \citeauthor{Colwell2018} is defined by the subsystems that are still functional. This may be appropriate for automated vehicles, but it is beyond the scope of this work.

In \textcite{Koopman2019}, the ODD is divided based on ODD considerations. For example, it focuses on specific ODD variants. This is distinct from the proposed work.

% ---------------------------------------------
%  Remote Driving + ODD + Trains (Tonk2021)
% ---------------------------------------------
\textcite{Tonk2021} consider the concept of restricted ODD (after \cite{Colwell2018} and $\mu$-ODD (after \cite{Koopman2019})) for remote-driven trains. 
The main aspect here is to transfer the concepts established for motor vehicles to the research area of trains.
The concepts are considered in a broader sense and risk minimal maneuvers are not mentioned in this work.

% ---------------------------------------------
%  ISO 23793-1
%  Intelligent transport systems Minimal risk manoeuvre (MRM) for automated driving Part 1: Framework, straight-stop andinlane stop
% ---------------------------------------------
ISO 23793-1~\cite{ISO23793-1:2024} describes a procedure how to perform a minimal risk maneuver with an ADS and the requirements for a vehicle system to execute a minimal risk maneuver.
It also explains how to test a minimal risk maneuver in an automated vehicle.
The standard does not address teleoperation and focuses on the use of an ADS.
In this article, we propose using a dedicated system to perform the minimal risk maneuver.
This is in stark contrast to ISO 23793-1, which states that ``[t]he MRM is not performed by a dedicated system but is a functionality that is implemented within an ADS''~\cite{ISO23793-1:2024}.

\section{Systematic Operational Design Domain Definition for Teleoperated Vehicles}
\label{sec: discussion}

This section will discuss the need for a \todd based on the capability of executing a minimal risk maneuver. 
The purpose of this section is to illustrate the problem of using teleoperation when the ADS cannot continue on its own.

\subsection{What is the problem?}

As mentioned in section \ref{sec: terminology}, the vehicle as a whole can be seen as a \textit{system of systems}, of which the ADS and teleoperation system are subsystems. 
These subsystems use the different capabilities of the overall vehicle system.

In this article, the focus is on the connection capability, which is used by both the ADS and the teleoperation system.
The difference between the ADS and the teleoperation system is that the ADS does not require the same level of quality for the connection capability as the teleoperation system.
If the connection fails or the quality degrades to an unusable level, the ADS is technically still able to execute its dynamic driving task.
However, if the same happens for the teleoperation system, the dynamic driving task cannot be continued by the teleoperation system without the help of an ADS or without posing an unreasonable risk.

In this article, we discuss the scenario in which teleoperation is applied as a fallback solution while the ADS is out of ODD and not able to operate. 
Then, if the connection quality degrades, the teleoperation system cannot use the help of an ADS because the ADS is not capable of working in this situation and there is no further backup solution.

In the research area of autonomous vehicles, the failure of a component's capability can most likely be addressed by redundancy options and dealt with at the system level in the ADS.
However, this is not the case for teleoperated vehicles because the connection depends on environmental structures, such as antennas, and the network's capacity utilization.
Although redundant network systems are used in teleoperated vehicles, they can neither guarantee nor directly influence mitigation of network failure because the aforementioned factors are beyond the vehicle's control.
For example, the connection depends on cellular towers, which the teleoperation system manufacturer cannot influence directly.

In this section, we will discuss what happens when the connection capability of a teleoperated vehicle with an onboard ADS degrades to a level that makes it no longer remotely drivable. 
This will underline that the degradation of the connection capability poses an unreasonable risk.
Simply speaking, we will ask the question: What happens if the connection is lost?

\subsection{Why is it a problem?}

From a safety perspective, it is important to consider what will happen when the teleoperated system's connection capability degrades, because the vehicle will no longer be controllable by the remote operator.
In this case, a system capable of reaching a minimal risk condition must stop the vehicle.
When an ADS is on board, such a system seems at first like a logical system for this maneuver.

However, when the teleoperation system is used out of the ODD of the ADS, the question arises as to whether it is capable of executing a minimal risk maneuver.

\subsubsection{Inside or outside of the ODD of the ADS?}

In the context of using an ADS for executing a minimal risk maneuver for a teleoperation system, the definition of the ODD is important. 
As mentioned in \ref{sec: terminology ODD} J3016 defines ODD as ``[o]perating conditions under which a given driving automation system or feature thereof is specifically designed to function, including, but not limited to, environmental, geographical, and time-of-day restrictions, and/or the requisite presence or absence of certain traffic or roadway characteristics ''~\cite[Clause~3.21]{J3016}.

When looking at the definition of SAE J3016, one could argue that, when only considering an ADS, the system could be either \textit{inside} or \textit{outside} the ODD of the ADS.
As can be seen, this definition makes sense for an ADS because it is written for an ADS and not for an ADS combined with teleoperation.

However, given the limitations of teleoperation with respect to attributes such as latency, jitter, packet loss, uplink, and downlink capabilities (e.g., as seen in \cite{Neumeier2019_feasability} and \cite{ Tener2023_Investigation}), it is questionable whether a boolean description of the ODD is meaningful for teleoperated vehicles.
We will discuss this in the next subsection.

What this article does \textit{not} want to focus on is the question of what happens when the quality of the connection capability lowers, but the remote operator is still able to operate the vehicle.
In this case, teleoperation can still be used, albeit in a degraded manner.
There are already detailed publication on this topic, as can be seen, for example in \cite{Neumeier2019_feasability, Neumeier2019_latency, Georg2020} and \cite{Brecht2024_Risk}.
This article focuses on what happens when the remote operator is unable to control the vehicle due to connection loss, posing a potential unreasonable risk.

\subsection{Ambiguity of a Boolean ODD description for an ADS}

% Nach oben gezogen für direktere positionierung
\begin{figure}[tbp]
\centering
\includegraphics[height=0.5\textheight, page=8]{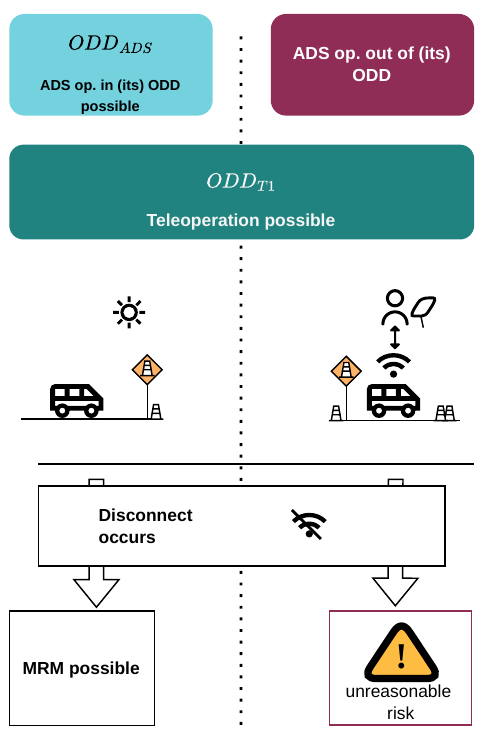}
\caption{
This is one of the current Boolean understandings of a \todd .
Here teleoperation is possible in both modes: inside and outside of the ODD of the ADS. The \todd $ODD_{T1}$ spans over both, the ODD of the ADS $ODD_{ADS}$ and outside of $ODD_{ADS}$.
When a disconnect occurs the minimal risk maneuver is only applicable while inside the ODD of the ADS.
When this happens outside of the ODD, the system poses an unreasonable risk.
This only provides a basic, demonstrative overview of the topic.
There may be other nuances.
}
\label{fig: from there to there 1}
\end{figure}

In case teleoperation is used to extend the ODD of the ADS, but an ADS is still onboard, there is the possibility for a degraded connection leading to the need of executing a minimal risk maneuver. 
There are two cases that need to be distinguished: (1) the quality degrades \textit{within} the ODD of the ADS, and (2) the quality degrades \textit{outside} the ODD of the ADS while driving the vehicle remotely.
An example of a Boolean understanding with two different ODDs can be seen in \cref{fig: from there to there 1}.
This shows a possible misunderstanding of the use of a teleoperation ODD. 
There is one ODD for the ADS, whereby a minimal-risk manoeuvre is still possible in the event of a disconnect, and one ODD for the teleoperation system.
In this example, teleoperation is possible within and outside the ADS's ODD.
However, outside of the ADS's ODD, the use of teleoperation poses an unreasonable risk, which is why we introduce the ODD understanding of \cref{fig: ODD_understanding_2}.

\subsubsection{Inside the ODD of the ADS?}

The first consideration is whether the vehicle can be driven remotely \textit{within} the ODD of the ADS.
In this case, we will use the construction site example from the introduction of this article (see \cref{fig: common_example}):
Teleoperation is used in this situation to remotely drive the vehicle through a construction zone.
If teleoperation is used inside the ODD of the ADS, one might ask why it is used at all.
Why not use the ADS to drive through the construction zone, when it is ``specifically designed to function''~\cite[Clause~3.21]{J3016} in this domain?

If the ADS cannot drive through this operational domain on its own, it is questionable or unclear if the ADS is actually within its ODD.
The question is whether, by definition, this ADS is out of its ODD if it cannot handle the situation alone.

Depending on the teleoperation function's specific technical implementation, this may vary from case to case. However, this shows that a discussion is needed to potentially define an ODD for ADS in combinantion with teleoperation.

\subsubsection{Outside the ODD of the ADS?}

The other consideration is when the construction zone mentioned in the introduction (see \cref{fig: common_example}) is \textit{outside} the ODD of the ADS.
This is consistent with the popular claim that teleoperation can be used to \textit{extend} the ODD of the ADS, when the ADS cannot be used in that operational domain (e.g., as seen in \cite{Brecht2024, Koskinen2024}).
Initially, this would solve many problems because the teleoperation system could be used for all scenarios in which an ADS does not know how to proceed. 
Everything that an ADS cannot do would be possible with the teleoperation function.
The construction zone example would be a standard use case for handling the aforementioned ODD uncertainty, because it is clear that the teleoperation function would extend the ADS's ODD.

However, it must also be clear that the vehicle is really outside the ODD of the ADS, because the ODD is being extended and the ADS cannot drive through the current operational domain on its own.

\subsubsection{Resulting consideration}

Both cases result in either (1) a unnecessary use of teleoperation or (2) a use of teleoperation without the backup of an ADS.

Using teleoperation when (1) completely inside the ODD of the ADS is a questionable task.
(As we show later a system only capable of executing a minimal risk maneuver would be better).
The uncertainty arises from the question of why invest resources in something unnecessary and sometimes problematic (e.g., with a remote operator, the network connection becomes an unpredictable and critical component).
It is unclear why teleoperation should be used in the ADS's ODD, but the connection capability would not be a problem because the ADS could take over.
Another factor that influences this, for \rd , the ADS must be in standby mode so it can activate immediately.
This requires a significant energy compromise of the whole system because the sensors and system must run in the background for limited benefits.

The second case (2) leads to the question: Why is it a problem if the ADS is not in its ODD when the teleoperation system needs assistance?
A critical issue that will be discussed in the following subsection is when the quality of the connection capability degrades to a level that is no longer drivable and a minimal risk maneuver must be executed.

\subsubsection{Relying on a non-capable system}
Normally, an automated vehicle will have an ADS on board that can function as a fallback. There should be the potential to use this for a minimal risk maneuver if the quality of the connection capability degrades.

In the construction zone example, it is unclear whether the ADS is capable of performing such a minimal risk maneuver.
The ADS is not ``specifically designed to function''~\cite[Clause~3.11]{J3016} in this scenario. 
That is why it needs help from a remote operator in the first place.
However, because the vehicle is outside of the ODD of the ADS, the execution of such a maneuver with the ADS is unpredictable, since by definition it is not designed to function in that ODD.

With this in mind, an ADS that cannot be used in that ODD is expected to serve as the backup system in case the connection to the operator station is lost.
This can be quite challenging, some would even call it a paradox, because it is difficult to rely on something that does not work properly in that ODD for such a critical maneuver.

Note: A similar problem arises when considering an unplanned ODD change while driving through the construction zone. For example, it starts to rain while the \ra is assisting the vehicle.
Again, it is unclear how the ADS should handle a minimal risk maneuver. The difference here is that the operator has still \textit{some} control.

\subsection{How to solve this problem?}

We suggest the implementation of a more specific \textit{tele\-operation ODD}, in which the vehicle system still has the defined capability to reach an adequate minimal risk condition. 

\begin{figure}[tbp]
\centering
\includegraphics[height=0.5\textheight, page=9]{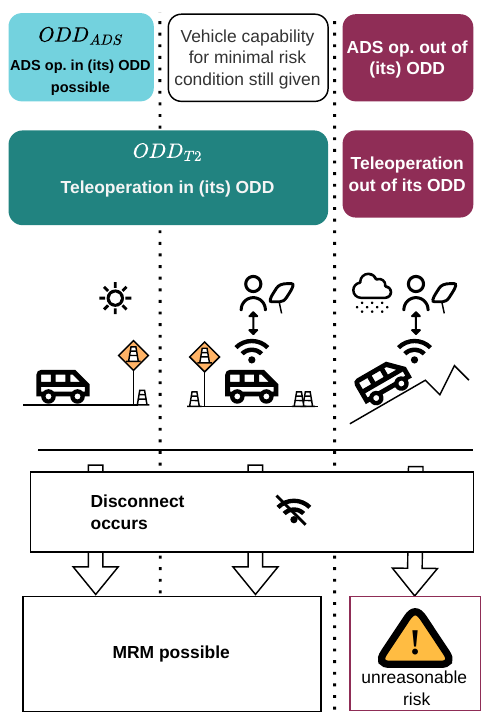}
\caption{
ODD description proposed in this article. A new mode has been added for the teleoperation function, allowing for driving a vehicle remotely within a defined ODD. This time the $ODD_{T2}$ of the teleoperation system spans over the ODD of the ADS and an ODD, where the vehicle capability for a minimal risk condition is still given.
However, there is still an area outside the defined ODD which has an unreasonably high risk of driving there remotely.
The capability of a minimal risk maneuver by the ADS is not guaranteed there, because it is outside of its ODD.
}
\label{fig: ODD_understanding_2}
\end{figure}

The proposed method is shown in \cref{fig: ODD_understanding_2} and will be further elaborated in \cref{sec: main idea}. 

\section{Proposed Idea}
\label{sec: main idea}

Our proposal is a \textit{\todd } based on the principle that teleoperation shall only be possible in the current operational domain if the vehicle system has the capability to perform a minimal risk maneuver.

In addition to the ODD of the ADS, there is a defined ODD of the teleoperation system. 
For this proposal, it does not matter whether the teleoperation system is operated via Remote Driving or Remote Assistance.

This section will at first describe all mode transitions shown in \cref{fig: ODD_unterstanding_1} and then discuss limitations and benefits of the proposal.

\subsection{Description of the mode transitions}

\begin{figure*}[t]
\centering
\includegraphics[width=\linewidth, page=5]{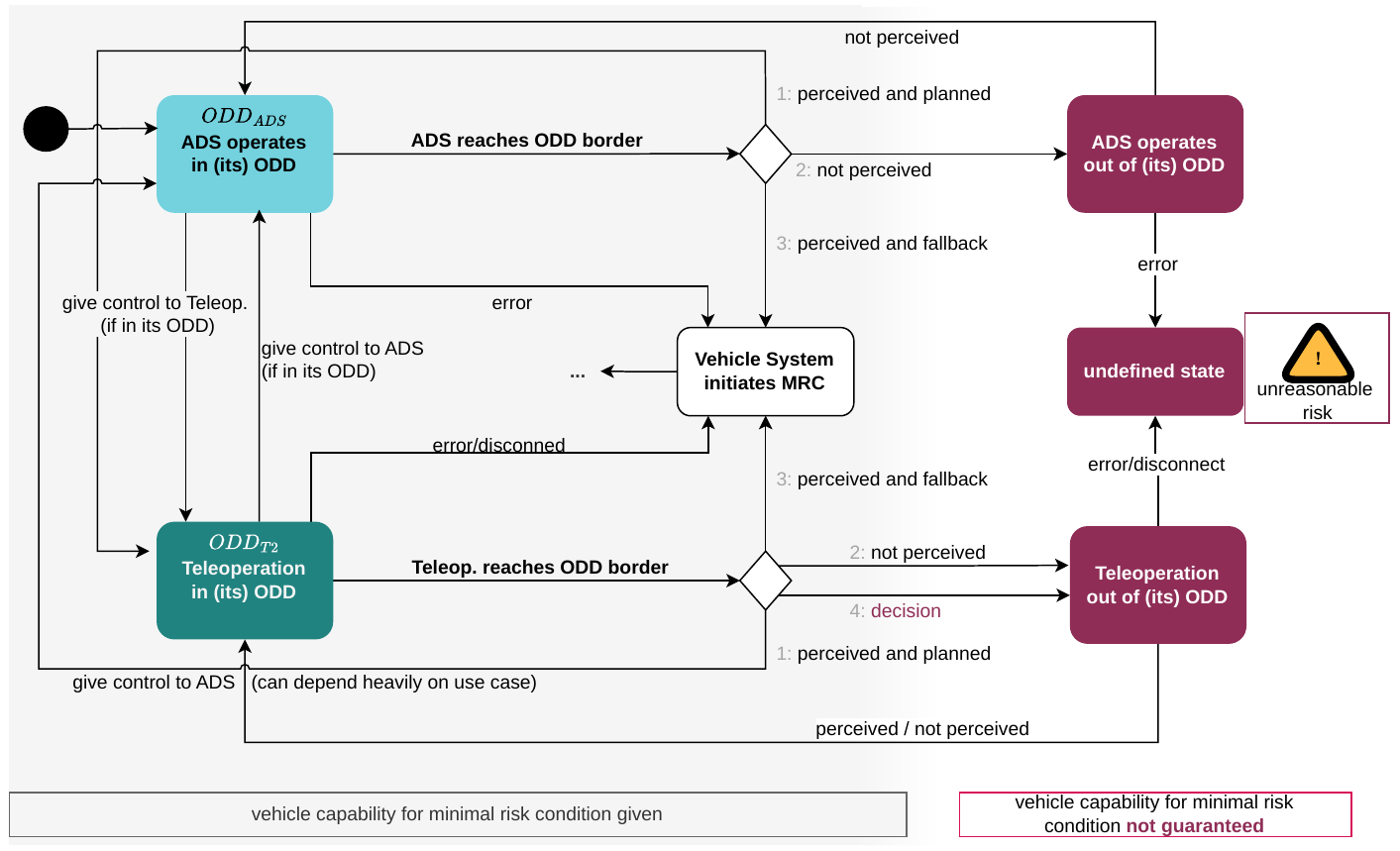}
\caption{
    Proposal for the integration of a \todd .
    Further elaboration is done in \cref{sec: main idea}.
}
\label{fig: ODD_unterstanding_1}
\end{figure*}

This subsection will present and discuss all modes and transitions between modes shown in \cref{fig: ODD_unterstanding_1}.
For clarity, the proposed method of a \todd based on the ability to perform a minimal risk maneuver does not include the red states (``ADS op. out of (its) ODD'', ``undefined state'' and ``Teleop. out of (its) ODD''). 
These states are included, to demonstrate the rationale behind this proposal.

\subsubsection{ODD ADS}
Initially, the vehicle is driven by an ADS. 
When an error occurs, the vehicle has the capability to transition to a minimal risk condition. 
The design of this condition is beyond the scope of this work.
For the purpose of this description, it is only important to note that the vehicle system has such a capability to initiate a minimal risk condition.
The manufacturer specified the ODD of the ADS to ensure that the vehicle system is capable of performing a minimal risk maneuver even if a potential error occurs.
As described by \textcite{Stolte2022}, the system is \textit{fail-safe}.

\subsubsection{Switch between ADS and Teleoperation}
When the current operational domain is inside the ODD of the ADS and also inside the \todd , then the ADS can give control to teleoperation, and the teleoperation system can give control back to the ADS.
This is standard behavior.

Reasons for switching from ADS to teleoperation could include a change in the operating environment, for example due to upcoming environmental changes or approaching rain.

\subsubsection{ADS reaching ODD border}
A more interesting case occurs when the ADS approaches its ODD limit.
This can happen for several reasons.
For example, it might start raining and the rain might intensify, but the ADS's ODD does not include heavy rain.
Another example is when the vehicle approaches a construction site.
Because construction sites are not included in the ADS's ODD, a remote operator could be called to drive or assist the vehicle through the site.

There are three options for the ADS when reaching an ODD border.
\begin{enumerate}
  \item Reaching the ODD border is perceived and planned. \newline
        As mentioned earlier, one option for reaching an ADS border could be that the remote operator takes control of the vehicle. 
        This would be a planned and perceived takeover.
  \item Reaching the ODD border is not perceived. \newline 
        Another possibility is that the ADS does not perceive being out of its ODD. 
        This could be the case if the current operational domain is not monitored correctly for all ODD borders. 
        In this case, the ADS would be operating out of its ODD.
  \item The ODD border is detected, and a fallback is \mbox{initiated}. \newline 
        The vehicle system will initiate a minimal risk maneuver when the ADS leaves its ODD, because this is a defined behavior. 
        It is important to note that the vehicle capability for a minimal risk maneuver is available, because the minimal risk maneuver is still planned within the ODD of the ADS. 
\end{enumerate}

\subsubsection{ADS out of its ODD}
In this state, the ADS performs the dynamic driving task out of its ODD. 
A hypothetical example of not perceiving the ODD is if the ADS could still execute the dynamic driving task, but could not execute a minimal risk maneuver.

The problem with this is that, in the event of a failure, the vehicle will be in an undefined state because it is not guaranteed that the system has the capability to initiate a minimal risk maneuver. 
There may be many other issues with this state that we will not address here, as the focus is on the \todd .
This is just an illustration of what happens when the transition out of the ODD is not detected.
This would, considering \textcite{Stolte2022}, be a \textit{fail-unsafe} state, because ``in the presence of a fault combination'' the system ''is not able to maintain a safe-state''.

Because an ADS can enter this state, it can also switch back to its ODD state.
When operating near an ODD boundary, there could be many unperceived switches between inside ODD and out of ODD.
The analysis of ODD boundary switches is not the focus of this article, but could be of interest for future work e.g., in the field of robust control engineering. 

If an error occurs while the ADS is out of its ODD, it would be in an undefined state, so it could be a matter of luck if the vehicle can execute a minimal risk maneuver.

Depending on the defined ODD border, a possible example for rapid switching would be concrete switching points. For instance, an ADS could be defined as being out of its ODD when there is a latency greater than \qty{250}{\milli\second} or when the humidity must be less than \qty{90}{\percent} and the current operating range is roughly around these values.

\subsubsection{Give control to Teleoperation}
When control is given to the remote operator, the vehicle system enters the proposed \todd .
Here, the remote operator controls the vehicle via either \ra or \rd .
If an error or disconnect occurs, the vehicle system still has the capability to initiate a minimal risk maneuver because the system is in an ODD designed for that purpose.
Considering \textcite{Stolte2022}, depending on whether the use of the teleoperation system is considered a degradation of the system, it is either in \textit{fail-degraded} or \textit{fail-safe} mode.
In either case, the system still has the capability to end in a safe state.

\subsubsection{Teleoperation ODD border}
Because the ODD of the ADS has an ODD border, the \todd also has one. 
When a teleoperation system reaches its ODD border, there are four options.
\begin{enumerate}
  \item Reaching the ODD border is perceived and planned. \newline 
  Similar to the ADS, the teleoperation system could give back control to the ADS. 
  An example of such a use case would be when the drive through the construction site is over and the ADS can take back control. 
  Depending on the use case, it is also possible to give back control due to environmental reasons, e.g., when it starts to snow.
  The limitation here, which might make other reasons more difficult, is that probably in most cases the \todd would include a ``bigger'' or different ODD than the ODD of the ADS. 
  
  \item Reaching the ODD border is not perceived. \newline
  Because the remote operator has the capability to drive outside of the \todd, it could be the case that reaching the ODD border is not perceived. 
  In this case, the teleoperation system would be driven outside of its ODD, which will be discussed in the next subsection.
  
  \item The ODD border is perceived and a fallback is initiated. 
  \newline If the ODD boundary is perceived by the monitoring system and no transition to ADS is planned,  
  a fallback can be initiated because the teleoperation system is still within its ODD and therefore the vehicle still has the capability to perform a minimal risk maneuver.
  
  \item Active decision to leave \todd .
  \newline  There could be an option to actively leave the \todd .
  While it would be technically possible to drive the teleoperated vehicle outside the proposed \todd , it is not recommended.
  If e.g., teleoperation is used everywhere an ADS cannot drive, then such a decision is made.
\end{enumerate}

\subsubsection{Teleoperation out of ODD}
\label{sec: teleop_out_of_ODD}
When a teleoperated vehicle operates outside of the \todd , the vehicle's capability to initiate a minimal risk condition is not guaranteed.
This means that if an error or disconnect occurs while driving in this mode, the vehicle system will be in an undefined state because there is no system in the vehicle designed to operate in the current operational domain. 
As discussed in the next subsection, this results in an unreasonable risk. 

From this state, it is possible to switch back to being in the \todd , either actively if it was decided to leave the ODD in the first place, or if it wasn't even perceived that the \todd was left.

\subsection{Limitations and Benefits}

This subsection discusses potential limitations and benefits of the proposed method.

\subsubsection{Unreasonable risk in undefined state}
What cannot be done with the proposed \todd is to have an unlimited ODD, because at some point the vehicle system will be outside of its ODD. 
The classic impossibility of a Level 5 vehicle problem applies here:
An nearly unlimited ODD implies that the system is designed to function under nearly every conceivable operating condition.
This includes obvious conditions such as rain or snow, but also non-trivial conditions such as floods or forest fires, which most human drivers often cannot even handle.
But even the known challenging operating conditions can lead to a large amount of specification, because the goal of extending the ODD now has to be specified to know in which cases the system should work to perform a minimal risk maneuver.

If the vehicle system is outside of its ODD, it is not clear how the vehicle will behave in the event of a fault in an undefined state. 

When an error or disconnect occurs, the vehicle would normally initiate a minimal risk maneuver. 
In this case, this is not possible because the vehicle's capability to perform a minimal risk maneuver is not guaranteed.
It may still be possible to reach a minimal risk condition, but it may also be possible that this would result in harm to either the passengers or other road users.
On the other hand, if for some reason the vehicle does not have the capability to perform a minimal risk \mbox{maneuver}, an obvious maneuver option is to brake without ADS, which will be discussed and excluded in \cref{sec: braking as mrm}.

For both the ADS and the teleoperation system, there is no system on board that is capable of handling such a situation.
For the ADS case, the system behavior is by definition undefined, because it would drive \textit{somehow} out of the operational domain for which it was designed.
For the teleoperation system, as long as there is no failure or disconnect, it is technically possible, but has an unreasonable risk, to drive outside the proposed \todd .
In terms of safety, having unreasonable risk within a driving mode is by definition of safety as ``absence of unreasonable risk''~\cite{ISO26262} \textit{unsafe}.
The vehicle is considering \textcite{Stolte2022} in a \textit{fail-unsafe} state then.
This makes it challenging that operating outside of the ODD can be allowed in a safety argumentation.

Even though driving in this mode is considered an unsafe method, it may be possible to do so when it increases the probability of mitigating harm due to other, external causes.
Examples of possible use cases that need further evaluation include:

\begin{itemize}
  \item Emergency situations when a passenger inside or outside the vehicle is in imminent danger.
  \item In natural disasters when there may be the potential to rescue people with remotely operated vehicles.
  \item In crisis situations such as war zones.
\end{itemize}

\subsubsection{No ODD extension possible}

Designing a \todd based on the vehicle's capability to reach a minimal risk condition places the focus on the system responsible for reaching the condition.
But the design of such a system restricts the ODD of the teleoperation system to this capability.
This, on the other hand, leads to the fact that it is not possible to extend the ODD of the ADS without limits (having an unlimited ODD is there not possible anymore) and it restricts the claim of teleoperation as an extension of the ADS to these capabilities. 
As a true extension, the system capable of the minimal risk maneuver must be able to handle difficult scenarios, because there is a reason why these scenarios were excluded from the ODD of the ADS.
Depending on the scenario, the dilemma of having unlimited real-world problems could also arise with this minimal risk maneuver.

On the other hand, limiting the task of this system to just reaching a minimal risk condition could also be something that helps both the development of this specific system capable of doing so, and the development of the ADS.

\subsubsection{Difference from network measurement proposals}
This proposal differs from other \todd considerations:
Instead of specifying the ODD based on areas where the connection to the operator station has to be reliable, this approach does not require prior measurements of the reliability and quality of the connection. 
The capability of reaching a minimal risk maneuver \textit{without} a connection is the main difference and the main safety factor here, because this capability ensures that a controllability is still given.
Measurements for such a geographical approach can be seen in \cite{Hans2023} and \cite{Neumeier2019_feasability} (Note: \cite{Neumeier2019_feasability} does not use the term ODD, but measures where teleoperation would be feasible based on connection parameters).

Such measurements and adaption of the teleoperation system can be integrated in this approach, as a minor degradation of the connection while driving remotely must not directly result in a minimal risk condition.
Such degradations can also be handled by adapting the driving style by driving slower, as proposed in e.g., \cite{Neumeier2019, Zhang2020}. 

The proposed idea of this article shifts the main safety requirement of a reliable connection to ODD considerations.
This brings the thinking process closer to more sophisticated automated driving processes.
From a safety perspective, this approach makes teleoperated driving more manageable because guaranteeing a \qty{100}{\percent} reliable connection is most likely impossible, even when using reliable connection areas, redundancy, and other methods.
The dependence on a reliable connection is critical in any case.
The idea that a connection for teleoperation cannot be guaranteed is also discussed in e.g., \cite{Tang2014, Hoffmann2022, WG_Teleoperation2024_en}.

Additionally, the proposal has the important advantage of enabling connections to undergo a quality management risk assessment rather than requiring an ASIL (Automotive Safety Integrity Level) rating, which would be difficult to assign.
However, this work is only a safety perspective on the issue.
From a mobility perspective, identifying areas that are suitable for teleoperation while teleoperation is possible to use (meaning in this work staying inside the \todd), remains useful and important.

\subsection{Aren't there already methods for this problem?}
Given the importance of achieving minimal risk conditions in teleoperated settings, one might expect existing solutions to already be in place.
While some ideas and requirements exist, there is no system whose sole function is to minimize risk in these settings.

In this section, we will discuss ideas that have been considered before and explore possible shortcomings that our approach could address.

\subsubsection{Necessity in law}
One could argue that it is necessary for an ADS to achieve an adequate minimal risk condition when driving with an \ra on board.
For example, the German Road Traffic Act (StVG)~\cite[section~2, paragraph~1e]{StVG} requires this:

\textit{``(2) Motor vehicles with an autonomous driving function must have technical equipment that is capable of [\dots] 7.) recognizing their system limits and, when a system limit is reached, when a technical malfunction occurs that impairs the performance of the autonomous driving function; or when the limits of the defined operating range are reached, of automatically setting the motor vehicle in a state that minimizes risk''} (free translation by the authors).

This paragraph describes three states in which the ADS should perform a minimal risk maneuver:
First, ``when a system limit is reached'', second ``when a technical malfunction occurs that impairs the performance of the autonomous driving function'', and third ``when the limits of the defined operating range are reached''.

Depending on how exactly a (1) system limit, (2) a technical malfunction, and (3) the limits of the defined operating ranges are interpreted, the ADS could be at one of these when it needs help from a remote operator.
First, all three states describe the transition from an ODD that the ADS can handle to an ``out of ODD-state''. 
With the described teleoperation function, the ADS is out of its ODD from the moment a problem occurs, making it much harder for the ADS to react. 
Additionally, when reaching the ODD boundary, the ADS shall be set to a state that minimizes risk. 
This would probably be a minimal risk maneuver rather than continuing of the trip with the help of a remote operator. 
It makes these options unclear for an ADS's request for help from a remote operator.

It is important to note that additional to the German Road Traffic Act, the Ordinance on the Approval and Operation of Autonomous Motor Vehicles ("Autonome-Fahrzeuge-Genehmigungs-und Betriebs-Verordnung - AFGBV"~\cite{AFGBV2022_en}) builds onto the rules described there, requiring a remote operator to assess the situation after a minimal risk maneuver.

\subsubsection{Braking or using another Level 2 function}
\label{sec: braking as mrm}

Technically, it is possible to consider another fallback mechanism that is not as \textit{intelligent} as the vehicle's ADS, such as braking or using some Level 2 functions.
However, an earlier article published by several of the authors~\cite{Brettin2025} (also mentioned in \cref{sec: related work}) shows that these mechanisms are not a solution and are critical.
In short, the article states that a Level 2 function is not designed to work without a supervisor, which is unavailable due to the disconnect. It also states that, in a common scenario where a human-driven vehicle is following a remotely operated vehicle, there is a high probability of a rear-end collision, even with only human-like decelerations.

It should be made clear how a minimal risk condition is considered in the work in this article.
One could argue that the lowest risk in a level 2 system is achieved by braking. 
However, such arguments can lead to risky behavior, as they could result in a solution that is implemented too quickly and is potentially unsafe. 
If something goes wrong, a jury or court will judge whether a less risky alternative could have been applied. 
A lower risk can be achieved by adding a dedicated system beforehand.

\subsubsection{Fallback trajectory}

In a special case, one could argue that a fallback trajectory can be planned by giving maneuver options to the technical supervisor while controlling the vehicle with \ra .
However, generating a fallback trajectory with the ADS might be difficult since the main reason for providing these trajectories to the operator is to enable supervision and stopping of the vehicle if necessary.
This is not possible when the vehicle is disconnected.

\subsubsection{Resulting consideration}

Examining the solutions implemented by various autonomous vehicle companies, it's unclear how the remote control functions in these situations.
Most of the time, the ADS seems to be active and asks what kind of ODD it is in.

With a dedicated system capable of achieving a minimal risk condition, ambiguity would be eliminated because the capability of achieving a minimal risk condition would still be present, even with a degraded connection.

\section{Use Case Example}
\label{sec: use case example}

\begin{figure}[tbp]
\centering
\includegraphics[height=0.5\textheight, page=11]{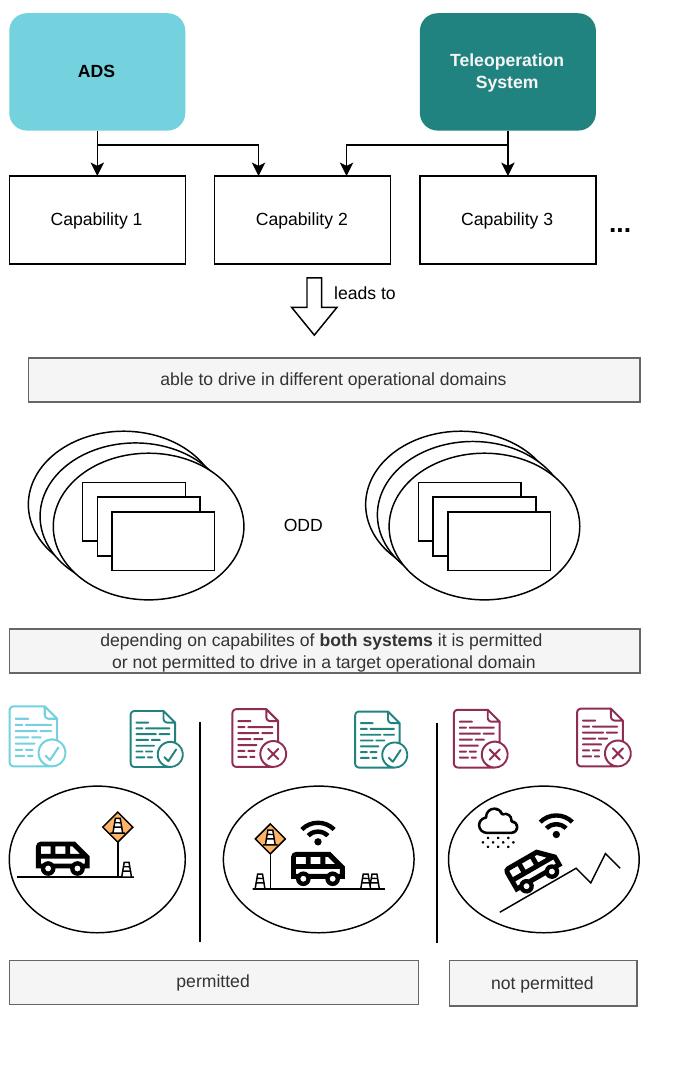}
\caption{
Use Case Example of the proposal shown in this article.
The ADS and Teleoperation Systems have different capabilities, leading to different potential Operational Design Domains in which they can drive. Depending on these capabilities, a regulatory authority can permit them to operate in one target domain but not another, such as in this case a snowy, hilly environment.
}
\label{fig: case study}
\end{figure}

For the purposes of this discussion, consider the proposal in the context of the use case shown in \Cref{fig: case study}.

As mentioned, the overall system includes an ADS and a teleoperation system with different capabilities. 
Some of the required capabilities are provided by both systems, while others are only available to one of them.
With the proposal of this article, we can define both ODDs in which the ADS and teleoperation have the capability to function. 
No unclear statement, with a teleoperation system that works everywhere is necessary.

With these definitions, a manufacturer can exactly define in which operational domain the vehicle system, including the ADS and the teleoperation system, is able to operate. 
For example, it could be the case that both systems are capable of handling highways with a speed limit of \qty{80}{\kilo\metre\per\hour}, which is potentially possible for both systems in Germany (see: \cite{StVFernLV2025, AFGBV2022}). 
However, only the teleoperation system can handle rain and perform a low-risk maneuver in these conditions. 
Neither system is able to handle snow. 
The ODD description could be more detailed, as e.g., seen in \cite{Schwalb2021, Irvine2021}, but the description should suffice for the example given in this section.

These definitions could then be submitted by the manufacturer to the different stakeholders to get an operating permit for an operational domain.

For example in Germany the document could be submitted to the Federal Motor Transport Authority (KBA ``Kraftfahrtbundesamt'') to get an operating permit and to the regional authority to get a permit to then drive in the related target operational domain, e.g., a highway in the city of Berlin.
There are also other relevant autorities like the German Technical Inspection Authority (TÜV ``Technischer Überwachungsverein'') or the German government itself, but going further into details into this process is out of scope for this work.

One possible outcome of the definitions in this example is that the system would be permitted to drive on this highway in summer, because it is capable of doing so.
In the described case, the ADS would drive in sunny conditions. If it starts to rain, the teleoperation system can be used to continue operating the vehicle. 
As there is a very low probability of snow in summer, the vehicle should be able to handle these situations. (Even if it starts to snow in summer, the ADS can still execute a minimal risk maneuver when it begins snowing.)
The advantage of this proposal is that, even if the ADS cannot drive in rainy conditions, the vehicle system as a whole is still in a safe and defined state because there is also an ODD definition for the teleoperation system that can operate in this domain.

In winter, however, as the probability of snow increases, the vehicle may not be permitted to drive in these conditions since both the ADS and teleoperation are outside their respective ODD. 
Using a teleoperation system as a backup here could lead to unsafe, unpredictable situations.

The advantage presented here is that both systems have a defined ODD. This means that, in any case, the vehicle is in a defined state and able to perform a minimal risk maneuver.

\section{Conclusion}
\label{sec: conclusion}

This work proposed the idea of a \todd based on reaching a minimal risk condition.
To this end, we presented potential mode switches and discussed the benefits and drawbacks of this approach.

This approach limits the commonly made promise to use teleoperation when ADS cannot continue on its own.
However, it eliminates the ambiguity of entering an undefined state when the ODD border is reached.
It also has the potential to shift the ASIL grade requirement for the connection to an on-vehicle component that does not require an active connection, making the safety argument for teleoperated vehicles more manageable.

Even if the initial problems seem specific, when one looks at the ODD definitions for a scenario that seems unclear, these scenarios are a common example of the use of \ra .
It is often unclear here whether the vehicle would be capable of ending up in a minimal risk condition.
The main purpose of this article is to demonstrate that this does not have to be unclear or hidden and to define an applicable state that does not rely on the connection of the teleoperated vehicle not to be lost.

We recognize that the use of such a system, capable of reaching a minimal risk condition, is in fact an SAE~Level~4 system, because there is no fallback driver to return control to if something unexpected happens.
Implementing such a solution is a challenging task, even though it is often argued that teleoperation should be used to mitigate the complications of an SAE Level 4 system. 
However, such a system can address one of the most critical safety issues by making the connectivity component of a teleoperated system manageable.

Future research should take a closer look at possible minimal risk maneuvers and what it means to be in a minimal risk condition for teleoperated driving.
This can even be of interest for future regulation, as the ambiguity of the usage of minimal risk conditions and the possibility of it being unsafe can be of interest here.

Studying the reactions of road users interacting with such a system would also be worthwhile.

\section{Acknowledgements}
We would like to thank Robert Graubohm for his valuable feedback on the contents of this article, and for discussing it with us.

We would further like to express our gratitude to Linda Block for proofreading the manuscript prior to submission.

\section*{Funding}
This work was partly supported by the German Federal Ministry for Economic Affairs and Energy within the project ``Automatisierter Transport zwischen Logistikzentren auf Schnellstraßen im Level~4 (ATLAS-L4)'' under grant number 19A21048H.
The funding source had no involvement in the study design, in the analysis and interpretation of the concepts presented, in the writing of the article, or in the decision to submit it for publication.

\section*{CRediT authorship contribution statement}
\textbf{Leon Johann Brettin:} Conceptualization, Methodology, Investigation, Visualization, Writing -- original draft, Writing -- review \& editing.
\textbf{Nayel Fabian Salem:} Investigation, Writing -- review \& editing.
\textbf{Ole Hans:} Writing -- review \& editing.
\textbf{Markus Maurer:} Supervision, Writing -- review \& editing.

\section*{Declaration of competing interest}
The authors declare the following financial interests/personal relationships which may be considered as potential competing interests:
Ole Hans reports a relationship with Vay Technology GmbH that includes: employment.
Vay Technology GmbH develops and operates remotely driven vehicles.
The other authors declare that they have no known competing financial interests or personal relationships that could have appeared to influence the work reported in this paper.

\section*{Declaration of generative AI and AI-assisted technologies in the writing process}

Because the authors of this article are not native English speakers, during the preparation of this work the authors used DeepL and Claude in order to translate and improve the language of the manuscript. After using these tools, the authors reviewed and edited the content as needed and take full responsibility for the content of the published article.

% Das hier war mal für das andere Konferenz-Format interessant
%\bibliography{IEEEabrv,references_offline}
%
\renewcommand*{\bibfont}{\footnotesize} 
%\addtolength{\textheight}{-1cm}

\printbibliography

\end{document}